\documentclass[twocolumn,showpacs]{revtex4}
\usepackage{amssymb}
\usepackage{amsmath}
\usepackage[dvips]{graphicx}

\begin{document}

\title{Hopping magneto-transport via nonzero orbital momentum states  and organic magnetoresistance}
\author{Alexandre S. Alexandrov$^{1,2}$, Valentin A. Dediu $^{3}$ and Victor V. Kabanov$^{1}$}
\affiliation{$^{1}$ Josef Stefan Institute,
1001 Ljubljana, Slovenia\\
$^{2}$ Department of Physics, Loughborough University,
Loughborough LE11 3TU, United Kingdom\\
$^{3}$ ISMN-CNR, Via Gobetti 101, 40129 Bologna, Italy}

\begin{abstract}
In hopping magnetoresistance of doped insulators, an applied
magnetic field shrinks the electron (hole)  s-wave function  of a
donor or an acceptor and this reduces the overlap between hopping
sites  resulting in the positive magnetoresistance quadratic in a
weak magnetic field, $B$.  We extend  the theory of hopping
magnetoresistance to states with nonzero orbital momenta. Different
from s-states, a weak magnetic field \emph{expands } the electron
(hole) wave functions with positive magnetic quantum numbers, $m>0$,
and shrinks the states with negative  $m$  in a wide region outside
the point defect. This together with a magnetic-field dependence of
injection/ionization rates  results in a \emph{negative} weak-field
magnetoresistance, which is linear in $B$  when  the orbital
degeneracy is lifted. The theory provides  a possible explanation of
a large low-field magnetoresistance in disordered $\pi$-conjugated
organic materials (OMAR).
\end{abstract}


\maketitle

As is well known the magnetoresistance (MR) in the hopping regime is
caused by a strong magnetic field dependence of the exponential
asymptotic of the bound state wave function at a remote distance
from a donor (or an acceptor) \cite{shklo}. In the case of the
Coulomb potential, when the wave function is spherically symmetric
in the absence of the magnetic field, it becomes cigar-shaped
squeezed in the transverse direction to the field \cite{shklo,miko}.
This leads to a significant decrease in the overlap of the
wave-function tails of two neighboring donors, and hence to a
significant increase of resistivity (positive MR). An exponential
positive MR in sufficiently strong magnetic fields is used to be a
hallmark of the hopping conduction. On the other hand, there is
anomalous (negative) MR observed in some hopping systems, for
instance in amorphous germanium and silicon. It has been attributed
to magnetic-field dependence of spin-flip transitions between sites
when some fraction of them has a frozen spin \cite{movag}, and/or to
an increase of the density of localised states due to the Zeeman
energy shift, $\mu_B B$ \cite{kam}. This negative MR is used to be
small (much less than 1\%) even in relatively high magnetic fields
of about 1 Tesla.

In   inorganic and organic insulators  lattice defects such as
vacancies,  interstitials, excess  atoms or ions and other
"impurities" often localise carriers with a finite momentum rather
than in the zero-momentum s-states. Here we extend the conventional
theory of magnetoresistance \cite{shklo} to hopping via non-zero
momentum orbitals. Quite remarkably this renders  a giant weak-field
magnetoresistance, which is negative. Moreover, if the orbital
degeneracy is lifted due to a broken time-reversal symmetry
\cite{landau} with or without net magnetization, the negative MR is
linear in $B$.

The  Schr\"odinger equation for the impurity-localised carrier
 wave function $\psi({\bf r})$ can be written in the integral form using the
 Green function, $G(\textbf{r},\textbf{r}^\prime;E)$, (GF) of the Bloch electron in a magnetic field,
\begin{equation}
\psi({\bf r})=-\int d{\bf
r}G(\textbf{r},\textbf{r}^\prime;E)V_{imp}({\bf r}^\prime)\psi({\bf
r}^\prime) \label{sch}
\end{equation}
where $E$ is the energy and $V_{imp}({\bf r})$ is the impurity
potential. We consider first a two-dimensional (2D) system, such as
a thin film in the magnetic field, $\overrightarrow{B}$
perpendicular to the surface of the film. Generally  GF is expressed
as a sum over wave-functions of the 2D Bloch electron in a rational
or irrational magnetic field  with the Hofstadter's butterfly
eigenvalues \cite{hof}. In a weak magnetic field with $\hbar
\omega_c=\hbar eB/m_b$ much smaller than the bandwidth, the
effective band-mass ($m_b$) approximation is sufficient, so that one
can use the 2D free-electron GF in  the magnetic field \cite{dod}
\begin{eqnarray}
&&G_{2D}(\overrightarrow{\rho},\overrightarrow{\rho}^\prime;E)={m_b\over
{2\pi \hbar ^2}}\exp\left[i {\rho \rho'
\sin(\phi'-\phi)\over{2l^2}}\right] \times \cr &&
e^{-(\overrightarrow{\rho}-\overrightarrow{\rho}^\prime)^2/4l^2}
\Gamma(a)U[a,1,(\overrightarrow{\rho}-\overrightarrow{\rho}^\prime)^2/2l^2],
\label{gf}
\end{eqnarray}
where  $\phi$ and $\phi^\prime$ are azimuth angles of
$\overrightarrow{\rho}$ and $\overrightarrow{\rho}^\prime$
respectively, $l=(\hbar/eB)^{1/2}$ is  the magnetic length,
$\Gamma(a)$ is the gamma-function, and $U(a,b,z)$ is the Tricomi's
confluent hypergeometric function well-behaved at infinity,
$z\rightarrow \infty$, for negative $E$ \cite{abram}. Here
$a=1/2-(E\mp \mu_BB)/\hbar \omega_c$, where $\mp$ corresponds to
spin up/down, respectively. Using Eqs.(\ref{sch}, \ref{gf}) one
finds  the wave function, $\psi({\bf r}) =F_m(\rho) \exp(i m \phi)$,
at $r_0\ll\rho \ll l^2/r_0$ as
\begin{equation}
 F_m (\rho,b)\propto \rho^{|m|} e^{-(\kappa \rho)^2 b/8} \Gamma(a) U\left[a,1,(\kappa \rho)^2 b/4\right],\label{sol}
\end{equation}
where $r_0$ is the radius of the impurity potential, $\kappa=(2m_b
\epsilon_0)^{1/2}/\hbar$ is the inverse localisation length of the
zero-field  state with the ionisation energy $\epsilon_0$, $b=B/B_0$
is the reduced magnetic field with $B_0=\hbar\kappa^2/2e$, and
$\rho$ is the distance from the impurity. While $F_m(\rho) \propto
\rho^{|m|} G(\rho,0;E)$ is strictly applied to any finite-range
$V_{imp}(\rho)$, it works for the infinite-range Coulomb potential
at large distances  as well \cite{shklo}. Neglecting a small
diamagnetic correction (quadratic in $b\ll 1$)
 yields $E=-\epsilon_0 + \hbar \omega_c m/2 \pm \mu_BB$ where $m=0, \pm 1,
\pm 2,...$ is the magnetic quantum number of the localised state,
so that
 \begin{equation}
a= {1\over{b}} +{1-m\over{2}}.
\end{equation}

To elucidate the magnetic field dependence of  the bound state we
expand the  solution, Eq.(\ref{sol}), in powers of $b$ making use of
the integral representation of $U(a,b,z)$ \cite{abram},
\begin{equation}
 U(a,1,z)= \Gamma(a)^{-1} \int_0^\infty e^{-zt}
 {t^{a-1}\over{(1+t)^{a}}}dt. \label{int}
\end{equation}
Replacing $t$ with $x=t/a$ and  $(1+ax)^a$ with  $ (ax)^a \exp[1/x-1/(2ax^2)+1/(3a^2x^3)-...]$ yields
\begin{eqnarray}
  {F_m(\rho,b)\over{\rho^{|m|}}} &\propto& 2K_0(\kappa \rho)+b (m-1){\kappa \rho \over{2}} K_1(\kappa \rho) + \cr
 && b {(\kappa \rho)^2\over{4}}[K_2(\kappa \rho)- K_0(\kappa \rho)] +{\emph{O}}(b^2),  \label{approx}
\end{eqnarray}
where $K_n(x)$ is the modified Bessel function. Eq.(\ref{approx}) is
applicable in weak magnetic fields, $b\ll 1$  far enough but not too
far from the point defect ($\kappa \rho \lesssim  1/b$). In a wide
region $1\ll \kappa \rho \lesssim 1/b$ one can use the asymptotic
\cite{abram} of $K_n(x)\approx
(\pi/2x)^{1/2}\exp(-x)[1+(4n^2-1)/8z]$ to get a leading
magnetic-field correction to the  wave function,
\begin{equation}
 {F_m(\rho,b)-F_m(\rho,0)\over{F_m(\rho,0)}}= m b {\kappa \rho\over{2}}.  \label{cor}
\end{equation}
Remarkably, the correction is \emph{linear} in B for any non-zero
$m$  and could be large, if the bound state is sufficiently shallow
and/or the hopping distance is large enough. It is positive for
positive $m$ and negative for negative $m$. The unusual expansion of
the wave function with the positive $m$  originates in the linear
lowering of the ionisation energy due to the orbital magnetic moment
in  weak magnetic fields.  On the contrary, the states with negative
$m$ shrink because their ionisation energy increases with the
magnetic filed. As shown below the state expansion/shrinking in the
region $\kappa \rho \lesssim 1/b$ causes a \emph{linear} MR in the
weak magnetic field, which is negative or positive depending on the
particular orbitals involved in the hopping.

\begin{figure}[tbp]
\begin{center}
\includegraphics[angle=-00,width=0.37\textwidth]{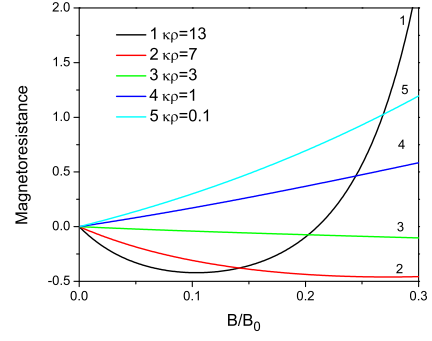}
\end{center}
\caption{(Color online) Hopping magnetoresistance in $m=2$ channel
as a function of the reduced magnetic field $B/B_0$   for a few
hopping distances. } \label{theory}
\end{figure}

In the case of  a finite-range impurity potentials, $V_{i,j}$, the
standard expression for the hopping integral \cite{shklo}, $t_{ij}=
<i\mid V_j\mid j> -<i\mid j><i\mid V_j\mid i>$, yields  $t_{ij}
\propto F_m(\rho)$, where $\rho$ is the distance between two hopping
sites $i$ and $j$, which is assumed to be much larger  than the
localisation length. The hopping conductance is proportional to the
hopping integral squared. Hence,  the magnetoresistance to the
hopping transport via particular $m$-orbitals is found as
\begin{equation}
MR_m\equiv{R(B)-R(0)\over{R(0)}}= {qK_0(\kappa
\rho)^2\over{pF_m(\rho,b)^2}} -1, \label{MR}
\end{equation}
where $p=\int_0^{\infty} dx x^{2m+1} K_0(x)^2$ and
$q=\int_0^{\infty} dx x^{2m+1} F_m(x,b)^2$
 accounts for  some weak field
dependence of the norm of the  wave function. The same expression is
also applied when hops take place between deep donor (acceptor)
levels and shallow levels with the emission or absorption of
phonons, as in the resistor network model of Ref.\cite{miller}.

Here and below we take $\rho$ as an average distance between defects
in the transverse direction to the field, so that $\rho \propto
N^{-1/2}$  in 2D  systems,  where $N$ is the density of defects
\cite{ref}. Fig.\ref{theory} represents hopping MR in the d-wave
orbital state with $m=2$ as a function of  the reduced magnetic
field for a few distances between defects. As one can see from the
figure the anomalous expansion of the bound state results in the
negative MR in weak magnetic fields, $b \lesssim (\kappa \rho)^{-1}
$ followed by the positive MR in stronger fields. The absolute value
of the negative MR could be in excess of 50 $\%$. The negative (for
$m>0$) and positive (for $m<0$) MRs are linear in $B$ at relatively
low $B$.

Let us now extend the theory to a 3D system. To incorporate the free
motion in the $z$ direction one can replace $E$ with $E-p_z^2/2m_b$
and perform the Fourier transformation in Eq.(\ref{gf}), $
G_{3D}(\textbf{r},\textbf{r}^\prime;E)= (1/2\pi
\hbar)\int_{-\infty}^{\infty} dp_z e^{ip_z (z^\prime -z)}
G_{2D}\left
(\overrightarrow{\rho},\overrightarrow{\rho}^\prime;E-p_z^2/2m_b\right)$.
Replacing $t$ in the integral representation of the confluent
hypergeometric function, Eq.(\ref{int}), with $t=1/(\exp(x)-1)$  and
integrating over $p_z$ one obtains
\begin{eqnarray}
&&G_{3D}(\textbf{r},0;E)={m_b \over {(2\pi)^{3/2}
\hbar^2}l}\int_{0}^{\infty} dx {e^{mx}\over{\sqrt{x} \sinh (x/2)}}
\times \cr &&\exp\left\{-\left [{(\kappa \rho)^2 b\over{8}}
+{x\over{b}}+{(\kappa z)^2 b\over{4x}}+{(\kappa \rho)^2 b\over{4
(e^x-1)}}\right]\right\}. \label{gf3D}
\end{eqnarray}
Expanding the exponent in the square brackets in Eq.(\ref{gf3D}) up
to the third power in $x$ and performing the integration by the
saddle-point method we finally obtain the asymptotic of the 3D wave
function, $\psi(\textbf{r},b) \propto
\rho^{|m|}G_{3D}(\textbf{r},0;E)$ at $1\ll \kappa r \lesssim 1/b$ as
\begin{equation}
\psi_m(\textbf{r},b)\propto \psi_m(\textbf{r},0) {\kappa r
b/2\over{\sinh(\kappa r b/2)}}\exp \left[{m \kappa r
b\over{2}}-{\kappa^3 \rho^2 r b^2\over{96}}\right]\label{3Dpsi}
\end{equation}
with $r^2=\rho^2+z^2$. For the s-wave bound state with $m=0$ this is
the textbook asymptotic \cite{miko,shklo} accounting for the
conventional positive MR quadratic in small $B$ . On the contrary,
for orbitals with nonzero orbital momentum the wave function,
Eq.(\ref{3Dpsi}), is linear in small $B$.

If there is no time-reversal symmetry breaking,   the states with
the opposite direction of the orbital angular momentum, $m$ and
$-m$, are degenerate, so that the linear term in the conductivity,
$\sigma=\sigma_m +\sigma_{-m}$ cancels,
\begin{equation}
\sigma (b)=\sigma(0)\left[{\kappa r b/2\over{\sinh(\kappa r
b/2)}}\right]^2 \cosh (m \kappa r b)e^{-\kappa^3 \rho^2 r b^2/48}.
\label{sigma}
\end{equation}
But even in this case  the hopping conductivity, $\sigma(b)$ first
\emph{increases} with the magnetic field (negative quadratic MR) and
only then decreases with $B$ (positive quadratic MR), if $\kappa
\rho^2/r < 24m^2-4$. Due to a large numerical factor ($24$), this
negative quadratic MR dominates in the whole region of realistic
impurity densities for any nonzero  $m$. Ions that carry a magnetic
moment break the time-reversal symmetry and split $m$ and $-m$
states. Such zero-field splitting (ZFS) gives preference to the
hopping via orbitals with a lower ionisation energy (positive $m$)
providing the \emph{negative linear} MR. In the extreme case of a
ferromagnet with a frozen magnetisation, magnetoresistance to
hopping via nonzero momentum orbitals should be highly anisotropic
changing from linear and negative in the field applied parallel to
the magnetisation to linear but positive in the opposite field, if
the  splitting due to exchange field is large enough compared with
the temperature.

\begin{figure}[tbp]
\begin{center}
\includegraphics[angle=-00,width=0.37\textwidth]{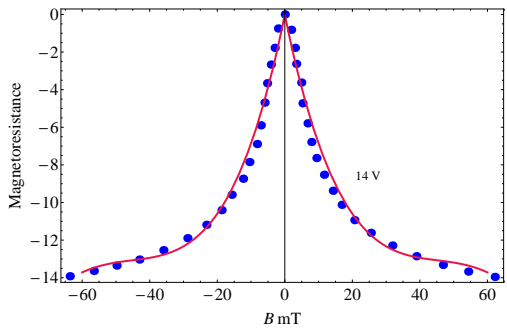}
\end{center}
\caption{(Color online) Negative OMAR (\%) at room temperature in
$ITO/PEDOT /Alq3/Ca$ device at  bias voltage $14$ V  (symbols,
Ref.\cite{mermer}) described with Eq.(\ref{final}) with $r=1.8$,
$B_s=64$ mT, and $B_m=40$ mT (solid line). } \label{exp1}
\end{figure}

\begin{figure}[tbp]
\begin{center}
\includegraphics[angle=-00,width=0.37\textwidth]{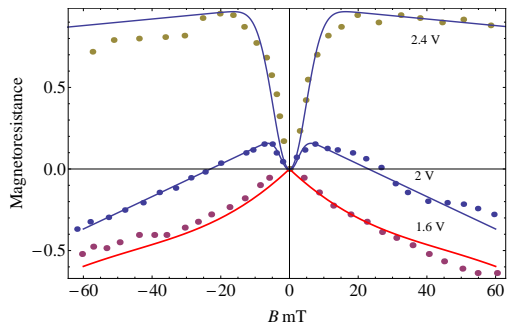}
\end{center}
\caption{(Color online) Transition from negative to positive  OMAR
(\%) at room temperature in  $ITO/PEDOT/pentacene/Ca$ device at
different bias voltages  (symbols, Ref.\cite{mermer})  described
with Eq.(\ref{final}) with $r=0.01$, $B_s=6.7$ mT,  $B_m=50$ T
(upper solid line), $r=0.0023$, $B_s=3.5$ mT, $B_m=10$ T (middle
solid line), and $r=0.01$, $B_s=63$ mT,  $B_m=5$ T (lower solid
line).} \label{exp2}
\end{figure}

At high electric fields the  conduction in organic and inorganic
insulators is often injection  and/or ionization limited where
carriers tunnel from extended states to a bound impurity level or
visa versa  under a potential barrier shaped by the electric field
\cite{arx}. The tunnelling ionization/injection rate, $W$ can be
described by a modified tunnelling ionization formula \cite{mod}
fitting the numerically calculated ionization rates of atoms and
ions over a large region of the electric field, $F$,
\begin{equation}
W_m \propto F^{-\beta} \exp\left(-{F_i\over{F}}-{\alpha
F\over{F_i}}\right), \label{ion}
\end{equation}
where $\alpha$ and $\beta$ are numerical constants, depending on a
particular ion, and $F_i\propto |\epsilon_0 -\hbar \omega_c
m/2|^{3/2}$ is the characterstic ionization electric field depending
on the magnetic filed in our case. Expanding $F_i$ in powers of $B$
yields
\begin{equation}
W_m (F,b) \approx W(F,0) e^{mb}, \label{ion2}
\end{equation}
where $W (F,0)\propto F^{-\beta} \exp\left(-F_{i0}/F-\alpha
F/F_{i0}\right)$ is the ionization/injection rate without the
magnetic field, $F_{i0} \propto \epsilon_0^{3/2}$, and  $b=B/B_{i0}$
is the reduced magnetic field with $B_{i0}= 4 B_0 F
F_{i0}/3(F_{i0}^2-\alpha F^2)$, which is a super-linear function of
the electric field. Like in the case of the hopping conduction the
injection/ionization conduction is a linear function of the weak
magnetic field, resulting in the linear negative MR, if the orbital
degeneracy is lifted.

 Finally, let us address  puzzling
experimental observations of negative and positive low-field MR in a
number of  organic materials \cite{mermer,bob,ngu}.  OMAR reaches
10$\%$ at fields on the order of only 10 mT, and can be either
positive or negative, depending on operating conditions. The Zeeman
energy does not account for the observed OMAR at ambient
temperatures since it  is too small, $\mu_B B \approx
 10$ mK,
at a field of 10 mT.  OMAR    in hole-only devices \cite{mermer,bob}
rules out exciton-based mechanisms as an explanation.  An
alternative model involving
 spin-dependent bipolaron formation in deep potential wells has been proposed
 \cite{bob}. Nevertheless the origin of OMAR is still debated.    Finding a convincing explanation of OMAR is important
for  understanding the basic transport mechanism of organic
insulators, which
 are used   in molecular spintronics
\cite{dediu,xiong}.

Here we  propose a more general model describing various classes of
materials characterized by hopping transport regime, which accounts
also for OMAR. We suggest that the hopping conductance could be a
combination of hoppings via conventional s-wave centers and via
non-zero angular momentum orbitals. There is experimental evidence
for paramagnetic centers and ZFS in polymers, in particular in
$Alq3$ \cite{para}. More recently (super)paramagnetic susceptibility
and ferromagnetic nanoclusters
 have been reported   in $Alq3$
\cite{FM}.  The conventional hopping magnetoconductivity is
described by the familiar exponential law \cite{shklo},
$\sigma_s=\sigma_{s0} \exp[(-B/B_s)^2]$, where $B_s$ depends on the
localisation radius and the density of s-wave centers, and
$\sigma_{s0}$ is the zero-field conductivity (see also
Eq.(\ref{sigma}) with $m=0$). The conductivity via non-zero angular
momentum orbitals split in zero field  is linear,
Eqs.(\ref{sigma},\ref{ion2}) and Fig.\ref{theory}, so one can
approximate it as $\sigma_m=\sigma_{m0} (1+B/B_m)$, where
$\sigma_{m0}$ and $B_m=B_0/(2|m|\kappa \rho)$ or $B_m=B_{i0}/|m|$ do
not depend on the magnetic field. As the result the combined
weak-field MR is described by the following simple expression,
\begin{equation}
MR={-B/B_m+r [1-\exp(-B^2/B_s^2)]
\over{1+B/B_m+r\exp(-B^2/B_s^2)}},\label{final}
\end{equation}
which can be readily compared with  experimental data for
sufficiently weak magnetic fields, $B\lesssim B_m$ (here
$r=\sigma_{s0}/\sigma_{m0}$). As one can see in Fig.\ref{exp1}, the
theory describes reasonably well the large negative OMAR  measured
at room temperature in $ITO/PEDOT /Alq3/Ca$ device at the bias
voltage $14$ V (and  other voltages in Fig.11 of Ref.\cite{mermer})
using $r, B_s, B_m$ as fitting parameters in the weak-field region,
$B\lesssim B_m$.  The low value of $B_m$ points to a dominating role
of hoppings via defects with rather shallow bound states (the
binding energy on the order of a few K) in this device. Some
empirical laws \cite{mermer}, in particular $\propto
-[B/(B+constant)]^2$ also gives accurate agreement, so that  the
good fit might be coincidental. However, as noticed in Ref.
\cite{mermer} simple fitting functions can fit  the data   only if
one stays away from the transition region between negative and
positive OMAR.  Remarkably, as illustrated in Fig.\ref{exp2},
simplified Eq.(\ref{final})   accounts for the cumbersome  MR also
in those organic structures, which show the transition  from
negative to positive MR. Compared with $ITO/PEDOT/Alq3/Ca$ device,
the $ITO/PEDOT/pentacene/Ca$ device in Fig. 14 of Ref.\cite{mermer}
has  deeper   bound states. The electric-field behaviour of $B_m$
and $r$ are in agreement with the injection/ionization conduction,
Eq.(\ref{ion2}) with a positive $\beta$.

In conclusion, we extended the conventional theory of  hopping
magnetoresistance  to hoppings via  non-zero orbital momentum
states. The asymptotic 2D and 3D solutions of the Schr\"odinger
equation show unusual linear expansion/shrinking of the bound state
with positive/negative magnetic quantum numbers far away but not too
far from the point defect in the magnetic field. Our theory accounts
for an extraordinary negative OMAR and for the transition to a more
ordinary positive MR in disordered $\pi$-conjugated organic
materials. Negative MR in some inorganic semiconductors  may  be
also reanalyzed in the framework of the theory.

This work was supported by  the European Union Framework Programme 7
(NMP3-SL-2011-263104-HINTS) and by the Royal Society (London).  We
thank Alexander Bratkovsky,  Alan Drew, Victor Khodel,  Yakov
Kopelevich, and  Ivan Naumov for enlightening  discussions.


\begin{thebibliography}{99}
\bibitem{ref} The resistance network model \cite{miller} and the percolation theory \cite{shklo} provide some  numerical corrections
to the doping dependence of MR, which can be accounted for by
redefining  $\rho$ in Eq.(\ref{MR}).
\end{thebibliography}
\end{document}